\documentclass[aps,prl,groupedaddress,showpacs,twocolumn,floatfix]{revtex4}
\usepackage{graphicx,bm,amssymb}
\usepackage{amsfonts}

\begin{document}
\title{Electric Dipole Spin Resonance for Heavy Holes in Quantum Dots} 
\author{Denis V. Bulaev and Daniel Loss}
\affiliation{Department of Physics and Astronomy, University of Basel, Klingelbergstrasse 82, CH-4056 Basel,
Switzerland}
\date{\today}

\begin{abstract} 
We propose and analyze a new method for manipulation of a heavy hole spin in a quantum dot.
Due to spin-orbit coupling between states with different orbital momenta and opposite spin orientations, an applied rf electric field induces transitions between spin-up and spin-down states. This scheme can be used for detection of heavy-hole spin resonance signals, for the control of the spin dynamics in two-dimensional systems, and for determining important parameters of heavy-holes such as the effective $g$-factor, mass, spin-orbit coupling constants, spin relaxation and decoherence times.
\end{abstract}

\pacs{67.57.Lm,76.60.Es,73.21.La} 

\maketitle

Spintronics, or spin-based electronics, is one of the fascinating and rapidly growing areas in  solid state physics and modern technology \cite{ALS}. Exploiting both the charge and spin degrees of freedom of carriers, it
offers wide opportunities for developing devices with unique functionalities. Operation of single charge or spin is the ultimate limit for such devices. Quantum dots (QD) have proven useful for this goal \cite{LV}, since experiments on readout and coherent manipulation of a single spin in QDs have already been performed successfully \cite{E2004,K2004,Petta,Blick}. Furthermore, due to supression of the spin-orbit interaction (SOI) in QDs \cite{AF,KN,GKL}, the electron spin in a QD has long relaxation times $T_1$ (up to hundreds of milliseconds) \cite{E2004,K2004,Zumbuhl}. The spin thus is an attractive candidate as carrier of quantum information \cite{LV}, if its state stays coherent over  sufficiently long times (described by the spin decoherence time $T_2$). At low magnetic fields, in III-V semiconductor QDs there is a rather strong hyperfine interaction between an electron spin and surrounding nuclear spins leading to a significant degradation of the spin coherence \cite{KLG,EN,CL}. Very recently, based on the idea of state narrowing via projective measurements \cite{CL}, several proposals have been made to limit the hyperfine-induced decoherence \cite{SBGI,KCL}, and, therefore, to increase $T_2$ times for electron spins in QDs, which currently range up to $\mu$s \cite{Petta,Blick}.

 Electron spin resonance (ESR) provides a powerful tool for coherent manipulation of spins, which is of great importance for spintronics \cite{Meisels}. By applying short resonant microwave pulses, an arbitrary superposition of spin-up and spin-down states is created. Rabi oscillations and spin-echo experiments are based on this approach \cite{Petta}. In such experiments, the ESR signal can be detected directly by measuring the absorption of radio-frequency (rf) power \cite{Meisels} via charge transport through a QD \cite{Blick,EL} or by using optical detection of magnetic resonance techniques \cite{Awschalom,Oliver}. Usually, ESR methods involve magnetic dipole transitions induced by an oscillating magnetic field. However, there has been a strong revival of interest in electric dipole spin resonance (EDSR) controlled by alternating \emph{electric} fields \cite{MR,Dobrowolska,RE,Kato,Mathias,Massoud} that provides the ability to manipulate and detect electron spins at the nanometer scale and may be useful for distinguishing different SOI mechanisms.

Recently, the idea to use heavy-hole (HH) spins (instead of electron ones) as carriers of quantum information has generated a lot of interest. On one hand, the hyperfine interaction of holes with lattice nuclei is suppressed, since the valence band has $p$ symmetry. On the another hand, the spin relaxation and decoherence for holes due to HH-phonon interaction could be comparable to or even longer than that for electrons in flat QDs \cite{BL}.

In this Letter, we propose a new method to study and control the spin dynamics of HHs in QDs via EDSR. We show that, due to SOI between states with different orbital momenta and opposite spin orientations, an applied rf electric field may induce transitions between spin-up and spin-down states and can be used for detection of a heavy-hole spin resonance signal and manipulation of HH spins in 2D systems. In contrast, rf magnetic fields do not affect the HH spin dynamics (in dipole approximation) and, therefore, are inefficient for manipulating and observing spin resonance and Rabi oscillations in HH 2D systems. 

We consider a single HH confined in a III--V semiconductor QD exposed to a static magnetic field $\mathbf{B}$ of arbitrary direction with respect to the [001] growth direction of a QD. Due to confinement along the growth direction, HH and LH subbands are split and, for thin enough QDs, SOI between LH and HH states can be taken into account perturbatively and, therefore, HHs can be treated as quasiparticles with pseudospin $\mathbf{S}$ aligned along the growth direction ($S_x=S_y=0$ and $S_z=3\sigma_z/2$, where $\sigma_z$ is the Pauli matrix) and with the following effective Hamiltonian \cite{BL},
\begin{equation}
\label{eq:H} H=\frac{1}{2m}(P_x^2+P_y^2)+U(x,y)+H_{\mathrm{SO}}-\frac12g_{\perp}\mu_\mathrm{B} B_\perp\sigma_z.
\end{equation}
Here $m$ is the effective HH mass, $\mathbf{P}=\mathbf{p}+|e|\mathbf{A(r)}/c$, $\mathbf{A(r)}=(-yB_\perp/2,xB_\perp/2,yB_x-xB_y)$, $\mathbf{B}=(B_x,B_y,B_\perp)$, $U(x,y)$ is the lateral confinement potential of a QD, $g_{\perp}$ is the component of the $g$-factor tensor along the growth direction, and 
\begin{equation}
\label{eq:Hso}
H_{\mathrm{SO}}=\beta P_-P_+P_-\sigma_++i\alpha P_-^3\sigma_++\frac{3\gamma_0\kappa\mu_\mathrm{B}}{m_0\Delta} B_{-}P_-^2\sigma_++h.c.
\end{equation}
is the SOI term of HHs consisting of three contributions:  the Dresselhaus term ($\beta$) \cite{BL}, the Rashba term ($\alpha$) \cite{Winkler2}, and the last term (due to SOI between LHs and HHs) combines two effects: orbital coupling via non-diagonal elements in the Luttinger--Kohn Hamiltonian ($\propto P_\pm^2$) and magnetic coupling via non-diagonal elements in the Zeeman term ($\propto B_\pm$)  \cite{Luttinger}.
Here, $\sigma_\pm=(\sigma_x\pm i\sigma_y)/2$, $P_\pm=P_x\pm iP_y$, $B_\pm=B_x\pm i B_y$, $\gamma_0$ and $\kappa$ are the Luttinger parameters \cite{Luttinger}, $m_0$ is the free electron mass, $\Delta$ is the splitting between LH and HH subbands,  and the SOI constants ($\alpha$ and $\beta$) depend on band parameters, confinement along the growth direction, and the energy spacing between LH and HH subbands $\Delta$ \cite{BL}. Note that we neglect the Zeeman splitting due to an in-plane magnetic field $\mathbf{B}_\parallel=(B_x,B_y)$ (because of strong anisotropy of HH $g$-factor: $g_\parallel\ll g_\perp$ \cite{KCPF}) as well as the orbital effect of the in-plane magnetic field, which is negligibly small for $B_\parallel\ll c\hbar/eh^2$ \cite{FAT}, where $h$ is the QD height. 

The Schr\"odinger equation for Hamiltonian (\ref{eq:H}) without SOI term ($H_{\mathrm{SO}}=0$) and with parabolic lateral confinement $[U(x,y)=m\omega_0^2(x^2+y^2)/2]$ has an exact solution \cite{FD}:
\begin{eqnarray}
\label{eq:Enms}
E_{n,m,s}&=& \hbar\Omega (n+1)+\hbar\omega_\mathrm{c}m/2-\hbar\omega_\mathrm{Z}s/3,\\
\label{eq:nms}
|n,m,s\rangle&=&c_{n,m}e^{im\varphi}\left( r/l\right)^{|m|}e^{-r^2/2l^2}L_{(n-|m|)/2}^{|m|}\left(  r^2/l^2 \right)\left|s\right\rangle,
\end{eqnarray}
where $(r,\varphi)$ are the polar coordinates, $n=0,1,2,\ldots$, the magnetic quantum number $m=-n,-n-2,\ldots,0(1),\ldots,n-2,n$  (which has the same parity as $n$), $s=\pm3/2$, $\Omega=\sqrt{\omega_0^2+\omega_\mathrm{c}^2/4}$, $\omega_\mathrm{Z}=g_\perp\mu_\mathrm{B}B_\perp/\hbar$ is the Zeeman frequency, $l=\sqrt{\hbar/m\Omega}$, $c_{n,m}= \sqrt{\left(\left(n-|m|\right)/2\right)!/\pi\left(\left(n+|m|\right)/2\right)!}/l$, and $L_n^m(x)$ is the generalized Laguerre polynomial.

Taking the SOI into account as a perturbation, we find that in first-order perturbation theory the two states corresponding to the Zeeman-split ground state level can be written as follows:
\begin{eqnarray}
\nonumber
|\pm\rangle&=& \left|0,0,\pm3/2\right\rangle
+i\beta^\pm_1\left|1,\pm1,\mp3/2\right\rangle+\beta^\pm_2\left|3,\pm1,\mp3/2 \right\rangle\\
&&+\alpha^\pm\left|3,\pm3,\mp3/2\right\rangle+\gamma^\pm B_\pm\left|2,\pm2,\mp3/2\right\rangle,
\label{eq:pm}
\end{eqnarray}
where 
$\beta^\pm_1=\beta(ml)^3\omega_\pm(\omega_{-}^2+\omega_+^2)/\hbar\omega^\pm_\mathrm{D}$, 
$\gamma^\pm=3\sqrt{2}\gamma_0\kappa\mu_\mathrm{B}(ml)^2\omega_\pm^2/m_0\Delta\hbar\omega_\parallel^\pm$,
and $\omega_\pm=\Omega\pm\omega_\mathrm{c}/2$
 ($\omega_\mathrm{D}^\pm$, $\omega_\parallel^\pm$, $\alpha^\pm$, and $\beta_2^\pm$ \cite{footnote1} are not relevant for the present discussion). 
From Eq.~(\ref{eq:pm}), it can be seen that $H_{\mathrm{SO}}$ leads to coupling of the two lowest states $|0,0,\pm3/2\rangle$ to the states with the opposite spin orientations and different orbital momenta. 
Due to this spin-orbit mixing of the HH states, the transitions between the states $|\pm\rangle$ with emission or absorption of an acoustic phonon become possible and this is the main source of the spin relaxation and decoherence of HHs (because the hyperfine interaction of holes with nuclei is suppressed) \cite{BL}. The spin relaxation between the states $|\pm\rangle$ due to Rashba spin-orbit (RSO) and Dresselhaus spin-orbit (DSO) terms is considered in detail in Ref.~\cite{BL}. Similarly, we can find the contribution $1/ T_1^\parallel$ of the third SOI term in Eq.~(\ref{eq:Hso}) to the spin relaxation rate $1/T_1$:
\begin{equation}
1/T_1^\parallel=\frac{B_\parallel^2\omega_\mathrm{Z}^5l^4}{2^8\pi\rho\hbar}\left( N_{\omega_\mathrm{Z}}+\frac{1}{2} \right)(\gamma^++\gamma^-)^2\sum_\alpha\frac{e^{-\omega_\mathrm{Z}^2l^2/2s_\alpha^2}}{s_\alpha^7}I^{(5)}
\label{eq:G1}
\end{equation}
(see Ref.~\cite{BL} for details). Note that in contrast to electrons \cite{GKL} there are no interference effects between DSO, RSO, and SOI due to LH-HH coupling and in-plane magnetic field, therefore, the total spin relaxation rate $1/T_1$ is the sum of rates $1/T_1=1/T_1^\mathrm{DSO}+1/T_1^\mathrm{RSO}+1/T_1^{\parallel}$. Note that in the limit of low temperatures ($k_BT\ll\hbar\omega_\mathrm{Z}$), $T_2=2T_1$ \cite{BL}. 

\begin{figure}
\includegraphics[clip=true,width=8.7 cm]{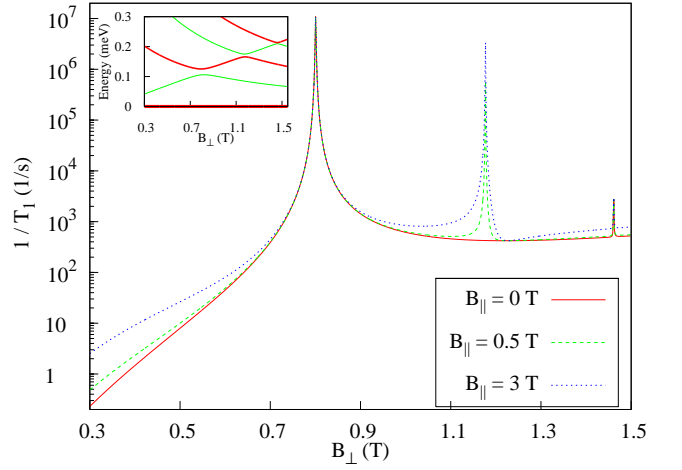}%
\caption{\label{fig:1} (color online). Heavy hole spin relaxation rate $1/T_1$ in a GaAs QD versus an applied perpendicular magnetic field $B_\perp$ (the height of a QD is $h=5\:$nm, the lateral size
$l_0=\sqrt{\hbar/m\omega_0}=40\:$nm, $\kappa=1.2$, $\gamma_0=2.5$, $g_\perp=2.5$ \cite{KCPF}, and the other parameters are given in Ref.~\cite{BL}). Inset: Energy differences of lowest excited levels with respect to the ground state $E_{0,0,+3/2}$.}
\end{figure}
In Fig.~\ref{fig:1} the total spin relaxation rate $1/T_1$ is plotted as a function of perpendicular magnetic field $B_\perp$. As can be seen from this figure and Eq.~(\ref{eq:G1}), the SOI due to in-plane magnetic field leads to an additional peak which is due to anticrossing \cite{BL,BL0} between the energy levels $E_{0,0,+3/2}$ and $E_{2,-2,-3/2}$ (see the second avoided crossing in the Inset). In the Inset, the first (third) avoided crossing resulting from DSO (RSO) coupling corresponds to the first (third) peak of the spin relaxation curve in Fig.~\ref{fig:1}.

Now let us consider methods for the  manipulation and detection of the HH spin in QDs. For electrons in 2D structures, an applied oscillating in-plane magnetic field couples spin-up and spin-down states via magnetic-dipole transitions and it is commonly used in ESR, Rabi oscillation, and spin echo experiments \cite{Petta}. From Eq.~(\ref{eq:pm}) it is easy to tell that magnetic-dipole transitions ($\Delta n=0$, $\Delta m=0$, and $\Delta s=\pm1$) are forbidden and, due to spin-orbit mixing of the states $|0,0,\pm3/2\rangle$ with $i\beta_1^\pm|1,\pm1,\mp3/2\rangle$, electric-dipole transitions ($\Delta n=\pm1$, $\Delta m=\pm1$, and $\Delta s=0$)  are most likely to occur.
Therefore, the HHs are affected by the oscillating electric field component and not by the magnetic one. 

We consider a circularly polarized electric field rotating in the XY-plane with frequency $\omega$: $\mathbf{E}(t)=E(\sin\omega t,-\cos\omega t,0)$. Therefore, the interaction of HHs with the electric field is described by the Hamiltonian $H^E(t)=(|e|E/m\omega)(\cos\omega tP_x+\sin\omega tP_y)$. The coupling between the states $|\pm\rangle$ is given by $\langle+|H^E(t)|-\rangle=H^E_{+-}=\left(  H^E_{-+}\right)^*=d_{\mathrm{SO}}Ee^{-i\omega t}$, where 
\begin{equation}
d_{\mathrm{SO}}=(|e|l/2\omega)(\beta_1^-\omega_-+\beta_1^+\omega_+)
\label{eq:d_SO}
\end{equation}
is an effective dipole moment of a HH depending on DSO coupling constants, perpendicular magnetic field $B_\perp$, lateral size of a QD, and frequency of an rf electric field ($\beta_1^\pm$ and $\omega_\pm$ are defined below Eq.~(\ref{eq:pm})). In the framework of the Bloch--Redfield theory \cite{Blum}, the master equation for the density matrix  in the rotating frame can be written as follows:
\begin{eqnarray}\nonumber
\dot{\rho}_{z}&=& 2(d_{\mathrm{SO}}E/\hbar)\rho_--(\rho_z-\rho_z^T)/T_1,\\
\label{eq:me}
\dot{\rho}_{+}&=&\delta_{\mathrm{rf}}\rho_{-}-\rho_+/T_2,\\
\nonumber
\dot{\rho}_{-}&=& -\delta_{\mathrm{rf}}\rho_{+}-2(d_{\mathrm{SO}}E/\hbar)\rho_z-\rho_-/T_2,
\end{eqnarray}
where $\rho_z=\rho_{++}-\rho_{--}$, $\rho_+=\rho_{+-}e^{i\omega t}+\rho_{-+}e^{-i\omega t}$, $\rho_-=i(\rho_{+-}e^{i\omega t}-\rho_{-+}e^{-i\omega t})$, $\rho_{nm}=\langle n|\rho|m\rangle$, $\rho_z^T=(W_{+-}-W_{-+})T_1$ is the equilibrium value of $\rho_z$ without rf field, $W_{nm}$ is the transition rate from state $m$ to state $n$, $T_1=1/(W_{+-}+W_{-+})$ is the spin relaxation time, $T_2=2T_1$ is the spin decoherence time \cite{BL}, $\delta_{\mathrm{rf}}=\omega_\mathrm{Z}-\delta\omega-\omega$ is the detuning of the rf field, and $\delta\omega$ is the resonance shift due to decoherence \cite{Blum}.

The coupling energy between a HH and an oscillating electric field is given by
\begin{equation}
\langle H^E(t)\rangle= \mathrm{Tr}(\rho H^E(t))=-\mathbf{d}_{\mathrm{SO}}\cdot\mathbf{E}(t),
\label{eq:H^E}
\end{equation}
where $\mathbf{d}_\mathrm{SO}=d_{\mathrm{SO}}(i\rho_{-+}-i\rho_{+-},\rho_{+-}+\rho_{-+},0)$ is the dipole moment of a HH. Therefore, the rf power $P= -d\langle H^E(t)\rangle/dt=-\omega d_{\mathrm{SO}}E\rho_-$ absorbed by a HH spin system in a stationary state is given by \cite{Abragam}
\begin{equation}
P= \frac{2\omega (d_{\mathrm{SO}}E)^2T_2\rho_z^T/\hbar}{1+\delta_{\mathrm{rf}}^2T_2^2+(2d_{\mathrm{SO}}E/\hbar)^2T_1T_2}.
\label{eq:P}
\end{equation}

\begin{figure}
\includegraphics[clip=true,width=8.7 cm]{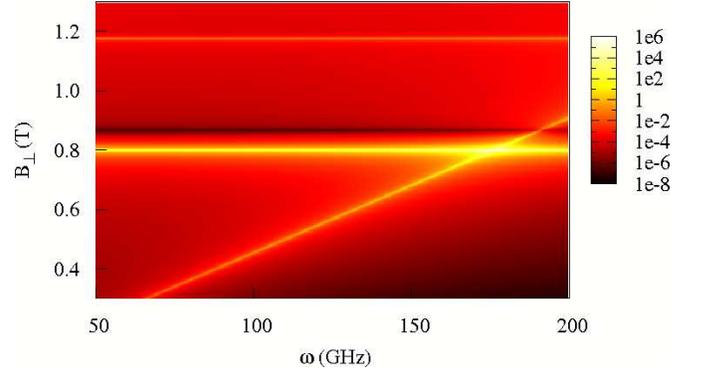}%
\caption{\label{fig:2} (color online). Absorbed power $P\:$(meV$/$s) as a function of perpendicular magnetic field $B_\perp$ and rf frequency $\omega$ ($T_2=2T_1$, $E=2.5\:$V$/$cm, $B_\parallel=1\:$T, and the other parameters are the same as those in Fig.~\ref{fig:1}).}
\end{figure}

In Fig.~\ref{fig:2}, the dependence of $P$ on a perpendicular magnetic field $B_\perp$ and frequency $\omega$ of the oscillating electric field is plotted. The rf power $P$ absorbed by the system has three resonances and one resonant dip. The first resonance occurs at zero detuning $\delta_{\mathrm{rf}}=0$. For the QD considered, the resonance shift $\delta\omega$ is negligible ($\delta\omega\approx1\:$kHz), therefore, the first resonance appears when the energy of rf radiation equals the Zeeman energy of HHs: $B_\perp^{\mathrm{r},1}=\hbar\omega/g_\perp\mu_\mathrm{B}$. The shape of this resonance (at certain $\omega$) is given by
$P\approx\hbar\omega\rho_z^T/2\hbar[1+\hbar^2\delta_{\mathrm{rf}}^2T_2/(2d_{\mathrm{SO}}E)^2T_1]$.
At low magnetic fields ($B<B^{\mathrm{r},2}_\perp$), $\delta\omega$ is the Lamb shift induced by zero-point fluctuations of acoustic phonons \cite{BL2007,footnote2}
\begin{equation}
\delta\omega\approx
\alpha^2C_{\mathrm{ph}}m^6l^5\omega_+^6/(3\omega_++\omega_\mathrm{Z})^2.
\label{eq:dw}
\end{equation}
The field dependence of $\delta\omega$
shows up in an effective $g_\perp$-factor which scales approximately as $B^{1/2}$.  
 
If the first and second resonances are well separated ($\omega\ll\omega_-$), then the absorbed power can be estimated as
\begin{equation}
P\approx 2\omega (d_{\mathrm{SO}}E)^2\rho_z^T/(\hbar \delta_{\mathrm{rf}}^2T_2),
\label{eq:P2}
\end{equation}
in the region of the second and third resonances and the resonant dip. The second resonance corresponds to anticrossing of the levels $E_{0,0,-3/2}$ and $E_{1,-1,3/2}$ (see the first avoided crossing in Inset of Fig.~\ref{fig:1}) at $\omega_-=\omega_\mathrm{Z}$ \cite{BL} (at $B_\perp^{\mathrm{r},2}=\hbar\omega_0/g_\perp\mu_\mathrm{B}\sqrt{1+2m_0/g_\perp m}$). At the anticrossing point, there is strong mixing of the spin-up and spin-down states and the dipole moment of a HH spin system is maximal $d_{\mathrm{SO}}^{\mathrm{max}}=|e|l\omega_\mathrm{Z}/2\omega$. Therefore, the height of the second resonance is given by $(el\omega_\mathrm{Z}E)^2/2\hbar\omega\delta_{\mathrm{rf}}^2 T_2$. The resonant dip appears at $B_\perp^\mathrm{d}=(\hbar\omega_0/2g_\perp\mu_\mathrm{B})\sqrt{2m_0/g_\perp m}$, which corresponds to $\beta_1^-\omega_-+\beta_1^+\omega_+=0$ and to zero dipole moment (see Eq.~(\ref{eq:d_SO})). The third resonance reflects the peak in the spin decoherence rate $T_2^{-1}$ due to applied in-plane magnetic field (see Fig.~\ref{fig:1}) at the second anticrossing point (the second avoided crossing in Inset of Fig.~\ref{fig:1}) at $2\omega_-=\omega_\mathrm{Z}$ ($B_\perp^{\mathrm{r},3}=4\hbar\omega_0/g_\perp\mu_\mathrm{B}\sqrt{1+4m_0/g_\perp m}$). From the positions of the resonances we can determine $g_\perp$, $m$, and $\omega_0$, from the shape and the height of those we can extract information about the SOI constants $\alpha$, $\beta$, and SOI strength due to in-plane magnetic field (which is proportional to $\gamma_0\kappa/\Delta$). Moreover, we can determine the dependence of the spin relaxation and decoherence times on $B_\perp$.

\begin{figure}
\includegraphics[clip=true,width=8.7 cm]{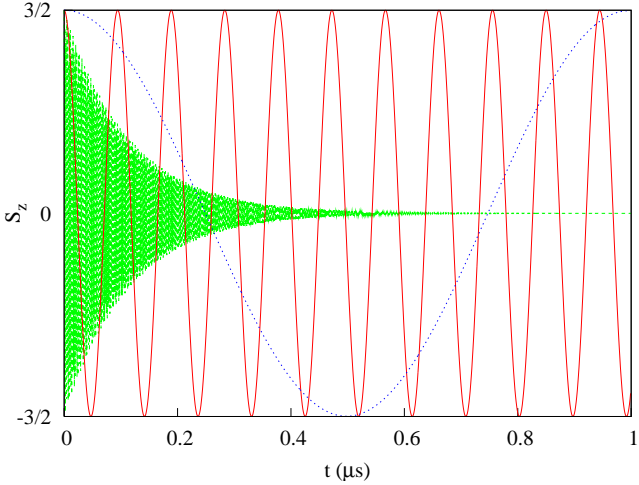}%
\caption{\label{fig:3} (color online). Rabi oscillations at three different values of the perpendicular magnetic field: $B_\perp=0.8\:$T (damped fast oscillations),  $B_\perp=0.865\:$T (dotted line), and $B_\perp=0.5\:$T (solid line). $B_\parallel=0$, $\delta_\textrm{rf}=0$, $E=1.5\:$V$/$cm, and the other parameters are the same as those in Fig.~\ref{fig:1}.}
\end{figure}
We now consider dynamics of the HH spin $\langle S_z\rangle=(3/2)\mathrm{Tr}(\rho\sigma_z)=(3/2)\rho_z$.
The Rabi oscillations at zero detuning ($\delta_\textrm{rf}=0$) of the spin are given by
\begin{eqnarray}\nonumber
\langle S_z \rangle &=& S_z^T+e^{-(T_1^{-1}+T_2^{-1})t/2}\left\{ \left( \frac32-S_z^T \right)\cos\omega_\mathrm{R}t\right.\\
&&\!\!\!\!\!\!\!\!\!\!\!\!\!\!\!\!\!\!\!\!\!\!\!\!\!\!\!\!\!\!\left.+\left[ \frac{(d_{\mathrm{SO}}E)^{2}T_2}{\hbar^2\omega_\mathrm{R}}S_z^T-\frac{T_1^{-1}-T_2^{-1}}{2\omega_\mathrm{R}}\left( \frac{3}{2}-S_z^T \right) \right]\sin\omega_\mathrm{R}t \right\},
\label{eq:S_z-t}
\end{eqnarray}
where $\omega_\mathrm{R}=\sqrt{(d_{\mathrm{SO}}E/\hbar)^2-(T_1^{-1}-T_2^{-1})^2/4}$ is  the Rabi frequency, and $S_z^T=(3/2)\rho_z^T/[1+(d_{\mathrm{SO}}E/\hbar)^2T_1T_2]$.

As can be seen from Fig.~\ref{fig:3}, the spin dynamics is essentially governed by the magnetic field $B_\perp$. The strong damping of Rabi oscillations is caused by fast HH spin relaxation at $B_\perp=0.8\:$T$\approx B_\perp^{\mathrm{r},2}$ (first peak in Fig.~\ref{fig:1}). A slight increase in $B_\perp$ drastically affects the dynamics of the HH spin (dotted line in Fig.~\ref{fig:3}): Rabi oscillations become stable due to much longer spin relaxation times and, at $B_\perp=0.865\:$T$\approx B_\perp^\mathrm{d}$, the effective dipole moment $d_{\mathrm{SO}}\approx0$ which substantially decreases the Rabi frequency $\omega_\mathrm{R}$.

In conclusion, we have introduced an efficient method for spin detection and manipulation of a HH in a QD. Furthermore, this method could be applied to determine important parameters of HHs in QDs (such as the effective $g$-factor, mass, SOI constants, spin relaxation and decoherence times), from the position and shape of resonances in the rf power absorbed by the system or from the amplitude evolution and the frequency of the Rabi oscillations.

\begin{acknowledgments} We thank L.~Vandersypen, B.~Coish, M.~Duckheim, and J.~Lehmann for useful discussions. We acknowledge
support from the Swiss NSF, NCCR Nanoscience, DARPA, ONR, and JST ICORP.
\end{acknowledgments}


\begin{thebibliography}{33}
\bibitem{footnote2}
$C_{\mathrm{ph}}=(3^4/2^5\sqrt{2\pi}\rho\hbar^3)[ (21/4) (eh_{14}/\varepsilon_\mathrm{s})^{2}\left( s^{-2}_\parallel+s_\perp^{-2} \right)-5(b/ls_\perp)^2 ]$. The parameters used here are determined in Ref.~\cite{BL}.
\end{thebibliography}
\end{document}